\documentclass[modern,longauthor]{aastex63} 
\accepted{2021 May 6}
\graphicspath{{./}{figures3.3/}}
\usepackage{amsmath}

\begin{document}

\title{The 3.3 $\mu$m Infrared Emission Feature: Observational and Laboratory Constraints on Its Carrier \\*[0.5in] }

\correspondingauthor{Alan Tokunaga}
\email{tokunagaa001@gmail.com}

\author[0000-0001-8136-9704]{Alan T. Tokunaga}
\affiliation{Institute for Astronomy \\
University of Hawaii \\
2680 Woodlawn Dr. \\
Honolulu, HI, 96822, USA \\
tokunagaa001@gmail.com}

\author[0000-0001-6150-1579]{Lawrence S. Bernstein}
\affiliation{Maine Molecular Sciences \\
14 Maine St., Suite 305B \\
Brunswick, ME 04011, USA \\
spectral1949@gmail.com}
 
\begin{abstract}

We examine the self-consistency of laboratory and observational data for potential carriers of the 3.3 $\mu$m infrared emission feature (IEF), a member of the ubiquitous family of strong interstellar IEFs at 3.3, 3.4, 6.2, 7.7, 8.6, 11.2, and 12.7 $\mu$m. Previous studies have shown that most Galactic sources (reflection nebulae, HII regions, and planetary nebulae) show 3.3 $\mu$m IEFs displaying similar central wavelengths, full widths at half maximum, and profiles.  Our study is focused on the band profile designated as Class A, the most prevalent of four classes of observed band profiles. 
In contrast to the observations, laboratory spectra for gas phase polycyclic aromatic hydrocarbons (PAHs), the widely assumed carriers of the IEFs, display central wavelength shifts, widths, and profiles that vary with temperature and PAH size. We present an extrapolation of the laboratory band shifts and widths for smaller PAHs ($\le$32 carbon atoms) to the larger PAHs ($>$50 carbon atoms) that are thought to be the IEF carriers. 
The extrapolation leads to tight constraints on the sizes of the putative PAH carriers.  Reconciling the observations with the implications of the laboratory spectra pose a significant challenge to the PAH and other IEF carrier hypotheses.
\end{abstract}

\keywords{Interstellar medium(847), Interstellar emissions(840)}

\section{Introduction} \label{sec:intro}

The family of strong interstellar infrared emission features (IEFs)\footnote{Over the decades, various names have been associated with the infrared emission features, such as the unidentified infrared bands, unidentified infrared emission features, aromatic infrared bands, and IEFs. We chose the acronym IEF in this paper as it is a neutral descriptor. }
at 3.3, 3.4, 6.2, 7.7, 8.6, 11.2, and 12.7 $\mu$m  have been studied since 1973 when the first emission band at 11.2 $\mu$m was discovered \citep{Gillett73}. Soon thereafter other strong IEFs were observed by \citet{Merrill75} and \citet{Russell77}.
Nearly 50 yr after their discovery, the carrier of the IEF bands has yet to be unambiguously identified.  
The goal of this paper is to establish constraints on the carrier of the 3.3 $\mu$m IEF band based on examining the self-consistency of previous observational and laboratory studies.
 
The most widely accepted explanation for the IEFs is that they arise from PAHs. The term “PAH hypothesis” was introduced by \citet{Allamandola89} to mean that “vibrations in individual, molecule-sized ($<$10 \AA) polycyclic aromatic hydrocarbons (PAHs) and PAH-like species were responsible for the bands”. \citet{Tielens08}, \citet{Peeters11}, \citet{Li20}, and \citet{Peeters21} review the PAH hypothesis.  The quantitative case for a stochastically heated  mixture of PAH molecules as the carrier of the IEFs was developed by \citet{Omont86}, \citet{Puget89}, and \citet{Allamandola89}.  The physics of infrared fluorescence is well known, and the conditions under which PAH molecules can survive in the interstellar medium (ISM) has been quantified.  In addition, the vibrational frequencies and band strengths for a large number of PAHs has been measured in the laboratory or estimated using computational chemistry.  

Since no individual PAH molecule has been found to provide a good match to all the IEF bands \citep[i.e.][]{Puget89} it is generally accepted that the IEF spectrum arises from a mixture of PAHs. 
Due to the destruction of small PAHs in the interstellar medium \citep{Allain96-1}, the interstellar PAHs are thought to be larger than any for which laboratory spectra exist, i.e. number of carbon atoms, $N_C$, is $>$50. 
It is further presumed that the mixture of many PAH molecules precludes any single PAH having sufficient abundance to be detected at UV or infrared wavelengths. 

The size range  of PAHs required to fit the IEF bands has been estimated based on the intensity ratio of the 3.3 and 11.2 $\mu$m IEFs and assuming these bands arise from the C--H stretch and C--H out-of-plane bending modes \citep{Pech02, Ricca12, Maragkoudakis20}.   A lower limit to the size range of the PAH molecule exists due to photodissociation and dehydrogenation of PAH by UV photons. \citet{Allain96-1, Allain96-2} found that PAHs smaller than approximately 50 carbon atoms are dissociated, as have more recent analyses by \citet{Montillaud13} and \citet{Andrews16}.
 
The term “PAH” in the astrophysical context includes molecular species that are not strictly polycyclic aromatic hydrocarbons as defined in chemistry (i.e., a chemical compound consisting of only carbon and hydrogen with multiple aromatic rings).  For example, the NASA Ames PAH IR Spectroscopic Database \citep[PAHdb;][]{Boersma14, Bauschlicher18, Mattioda20} includes PAH-like molecules with nitrogen and  oxygen, PAHs with five- and seven-membered carbon rings, and PAHs with aliphatic side groups.  The fullerenes C$_{60}$, and C$_{70}$ are also included.  \citet{Hudgins05} find that substituting one of the carbon atoms with nitrogen allows an improved fit to the wavelength of the 6.2 $\mu$m IEF.  The PAHdb also includes PAH cations and anions, as the charge state of the PAHs also plays a key role in fitting of the IEFs \citep{Boersma15, Boersma18}.  The continuum and {\em plateau emission} features are attributed to PAH clusters or very small grains \citep{Berne07, Pilleri12, Peeters17}. 
 
In addition to PAHs, there have been other proposed carriers of the IEFs.  These include hydrogenated amorphous carbon \citep[HACs;][]{Scott97, Gadallah13}, quenched carbonaceous composite \citep[QCC;][]{Sakata90}, coal materials \citep{Papoular89}, carbon nanoparticles \citep{Hu08}, and mixed aromatic-aliphatic organic nanoparticles \citep[MAONs;][]{Kwok13}. A summary of proposed carriers for the IEFs is given by \citep{Yang17-1}. The major limitations of solid-state candidates are that it is not yet possible to study very small particles of the solid-state materials under interstellar-like conditions and the laboratory analogs do not definitively fit all the IEFs.  The IEFs must also occur in particles of size 1--2 nm or less through stochastic heating and this is not yet possible to simulate this in the laboratory.  

There are four classes of IEFs, denoted by A, B, C, and D \citep{vanDiedenhoven04, Sloan14}.
This paper focuses on the Class A 3.3 $\mu$m IEF because (1) it is the most common type in the galaxy; (2) its assignment is well established as originating from aromatic C–H stretching fundamental (CH(S)) band, and (3) its profile is free from contamination from other fundamental bands and from combination, overtone, and hot bands. Nevertheless, the spectroscopy of the PAH CH(S) band is complicated by the presence of multiple Fermi resonances, which are manifested as intricate polyad structures \citep{Mackie15}.  The polyad structures are sensitive to the internal temperature and are molecule specific; hence, if the 3.3 $\mu$m IEF is attributed to PAHs, then one expects to see 3.3 $\mu$m IEF profile variations for different excitation environments.  Substantial variations of the CH(S) band profile with excitation energy for different PAHs are evident in the theoretical spectra presented by \citet{Mackie18} and the laboratory spectra obtained by \citet{Joblin94,Joblin95-1}.  

The paper is organized as follows.  In Section 2, we present a set of high spectral resolution and high signal-to-noise spectra of the 3.3 $\mu$m IEF that were obtained from Infrared Space Observatory (ISO) short wavelength spectrometer \citep[SWS;][]{Leech03}.  As noted previously and emphasized in Section 2, we show that the spectral profiles for the Class A 3.3 $\mu$m IEF are nearly identical for diverse sources. In Section 3 we present an analysis of the laboratory PAH spectra that exhibit size and temperature-dependent band shifts and widths for the CH(S) band.  We extrapolate the laboratory spectra to the larger PAHs that are thought to give rise to the IEFs in order to compare to the observations.  In Section 4 we discuss the implications of our findings on the carrier and its excitation by uv photons. 
%% =============================================== Sec. 2
\section{The 3.3 $\mu$\lowercase{m} IEF profile} \label{sec:3.3 profile}

%% =============================================== Sec. 2.1
%% \subsection{Emission profile, central wavelength, and full width at half maximum (FWHM)}  \label{sec:profile}

Observations of the 3.3 $\mu$m IEF show that most sources have similar IEF profiles and central wavelengths \citep{Muizon90, Tokunaga91, vanDiedenhoven04, Peeters11, Smith20}.  \citet{vanDiedenhoven04} classified the 3.3 $\mu$m IEF profiles into Classes A, B1, and B2 using the spectra of galactic and extragalactic sources obtained with the ISO SWS \citep[][]{Leech03}.  The different classes reflect differences in the center wavelengths and widths of the observed profiles with Class A being the most common class in galactic and extragalactic sources.

%% Figure 1
%% The "ht!" tells LaTeX to put the figure "here" first, at the "top" next
%% and to override the normal way of calculating a float position
\begin{figure*}[!ht]
\plotone{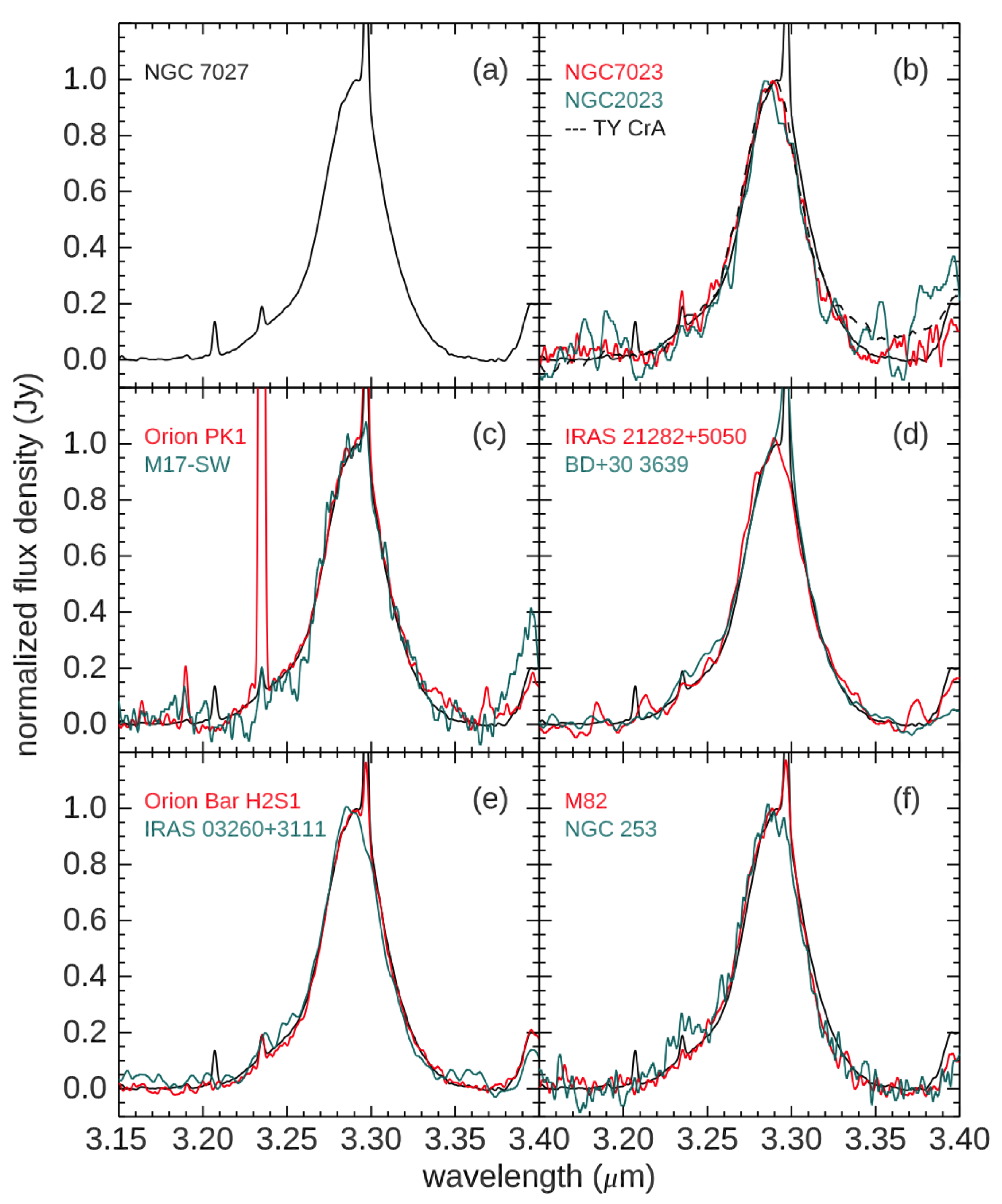}
%% original figure "Fig 1 axes fixed cut final2.pdf"
\caption{Mosaic of ISO SWS spectra obtained from the ISO spectral archive of \citet{Sloan03}. The spectra resolution, ISO TDT number, and physical parameters of the objects are given in Table 1.  Note that the spectra in panels (b)-(f) also include the NGC 7027 profile (black line) as a common point of reference.  
\label{fig:mosaic}}
\end{figure*}

Figure 1 shows a comparison of the 3.3 $\mu$m Class A profiles of galactic and extragalactic sources that were obtained from the ISO spectral archive of \citet{Sloan03}.  We compare the profiles to the profile of a weighted average of the NGC 7027 spectra listed in Table 1 because this object has the best set of high signal-to-noise spectra. The weights used were proportional to the signal to noise.  These spectra have had the continuum and 3.4 $\mu$m plateau feature removed as described by \citet{vanDiedenhoven04}. We did spot checking of the \citet{Sloan03} archive to \citet[][and E. Peeters 2021, private communication]{vanDiedenhoven04} and to the highly processed data products in the ISO Data Archive \citep{Frieswijk07} and found the 3.3 $\mu$m profile to be consistent among all three data sets.

Table 1 summarizes the properties of the objects. These were selected from the list of objects in \citet{vanDiedenhoven04} and with the addition of M17 SW \citep{Verstraete01} and TY Cra \citep{Seok17}.  All of the objects selected are among the highest signal to noise available in the ISO SWS archive and have a sufficiently high spectral resolution to clearly differentiate the  Pfund-$\delta$ hydrogen line at 3.297 $\mu$m from the 3.3 $\mu$m IEF.  Table 1 also shows the SWS TDT number, which identifies the specific SWS spectrum that is shown in Figure 1.

\begin{deluxetable*}{lcccccc}
%% \tablenum{1}
\tablecaption{List of Sources (Class A)  \label{tab:sources}}
\tablewidth{0pt}
\tablehead{
\colhead{Object} & \colhead{Type} & \colhead{TDT} & \colhead{{\em R} ($\lambda / \Delta \lambda$)} & \colhead{ {\em T}$_{\rm eff}$(K)/10$^3$ } & \colhead{$G_0$/10$^4$} & \colhead{References} 
}
\decimalcolnumbers
\startdata
NGC 7027        & PN  & 02401183, 33800505 & 1000 & 200 & 60 & 1, 2, 3  \\
                &     & 55800537 &      &     &      &     \\
IRAS 21282+5050 & PN  & 15901777 & 500  & 28  & 37   & 4, 5 \\
BD +30 3639     & PN  & 86500540 & 500  & 55  & 10   & 6, 7 \\
Orion Bar H2 S1  & \ion{H}{2} & 69501806 & 1000 & 37  & 4.2  & 8   \\
Orion PK1        & \ion{H}{2} & 68701515 & 1000 & - & -  & - \\
M17 SW	         & \ion{H}{2} & 32900866 & 1000 & 45  & 1.2  & 8   \\
NGC 7023 NW PDR  & RN  & 20700801 & 1000 & 15  & 0.26 & 9   \\
NGC 2023         & RN  & 65602309 & 500  & 23  & 0.12 & 8   \\
IRAS 03260+3111 & RN  & 65902719 & 500  & 13  & 2    & 7, 10   \\
TY CrA          & RN  & 33400603, 71502003 & 500 & 12 & 1 & 11, 12   \\
M82	            & EG  & 11600319 & 1000 & - & -  & -   \\
NGC 253	        & EG  & 24701422 & 1000 & - & -  & -   \\
\enddata
\tablecomments{PN = Planetary nebula; \ion{H}{2} = \ion{H}{2} region, RN = reflection nebula, EG = extragalactic.\\
TDT = ISO SWS target dedicated time identification number.\\
{\em T}$_{\rm eff}$ is the effective temperature of the exciting star.
$G_0$ is the spectrally integrated UV radiation flux from 5.17 to 13.6 eV  expressed as a ratio  to that in the solar neighborhood.  $G_0 = 1$ corresponds to 1.6 $\times$ 10$^{-3}$ erg cm$^{-2}$ s$^{-1}$ as calculated by Habing (1968).\\
References.  
(1) \citet{Latter00}.  (2) \citet{Justtanont97}.  (3) \citet{Stock16}.  (4) \citet{Leuenhagen98}. (5) \citet{Pech02}. (6) \citet{Crowther06}.  (7) \citet{Peeters02}.  (8) \citet{Verstraete01}; assumed a blackbody for the exciting star and so the $G_0$ estimates should be viewed with caution.  (9) \citet{Montillaud13}.  (10) \citet{Seok17}.   (11) \citet{Casey98}  (12) $G_0$ computed using eq. (1) in \citet{Stock16}.  
}
\end{deluxetable*}

We see that the profiles are very similar even though the objects have a wide range of physical conditions and evolutionary history, as previously noted by \citet{vanDiedenhoven04} and others.  Figure 1 demonstrates that the entire band profile is invariant, including the profile asymmetry and shape of the wings which are profile characteristics independent of and in addition to the central wavelength and FWHM.

The largest, although quite small, profile variability is seen for the reflection nebulae spectra NGC 7023, NGC 2023, and TY CrA (see Figure 1b).  The FWHM of these objects is slightly less with the long-wavelength side of the profile at a slightly shorter wavelength compared to NGC 7027.  
The average center wavelength and FWHM of NGC 7027, NGC 7023, Orion Bar H2 S1, Orion PK1, Orion PK2, NGC 2023, BD+30 3639, IRAS 21282+5050, and IRAS 03260+3111 were measured individually and averaged to get a center wavelength of 3.2887 $\pm$ 0.0009\ $\mu$m (3040.7 $\pm$ 0.8\ {\rm cm}$^{-1}$) and an FWHM of 0.0407 $\pm$ 0.0012\ $\mu$m (37.6 $\pm$ 1.1\  {\rm cm}$^{-1}$), respectively. The uncertainties given are the 1$\sigma$ standard deviation.  The center wavelength and FWHM agree with the values given by \citet{vanDiedenhoven04}.

%% The "ht!" tells LaTeX to put the figure "here" first, at the "top" next
%% and to override the normal way of calculating a float position
%% =========================== Figure 2 
\begin{figure}[ht]
%% \epsscale{1.2}
\plotone{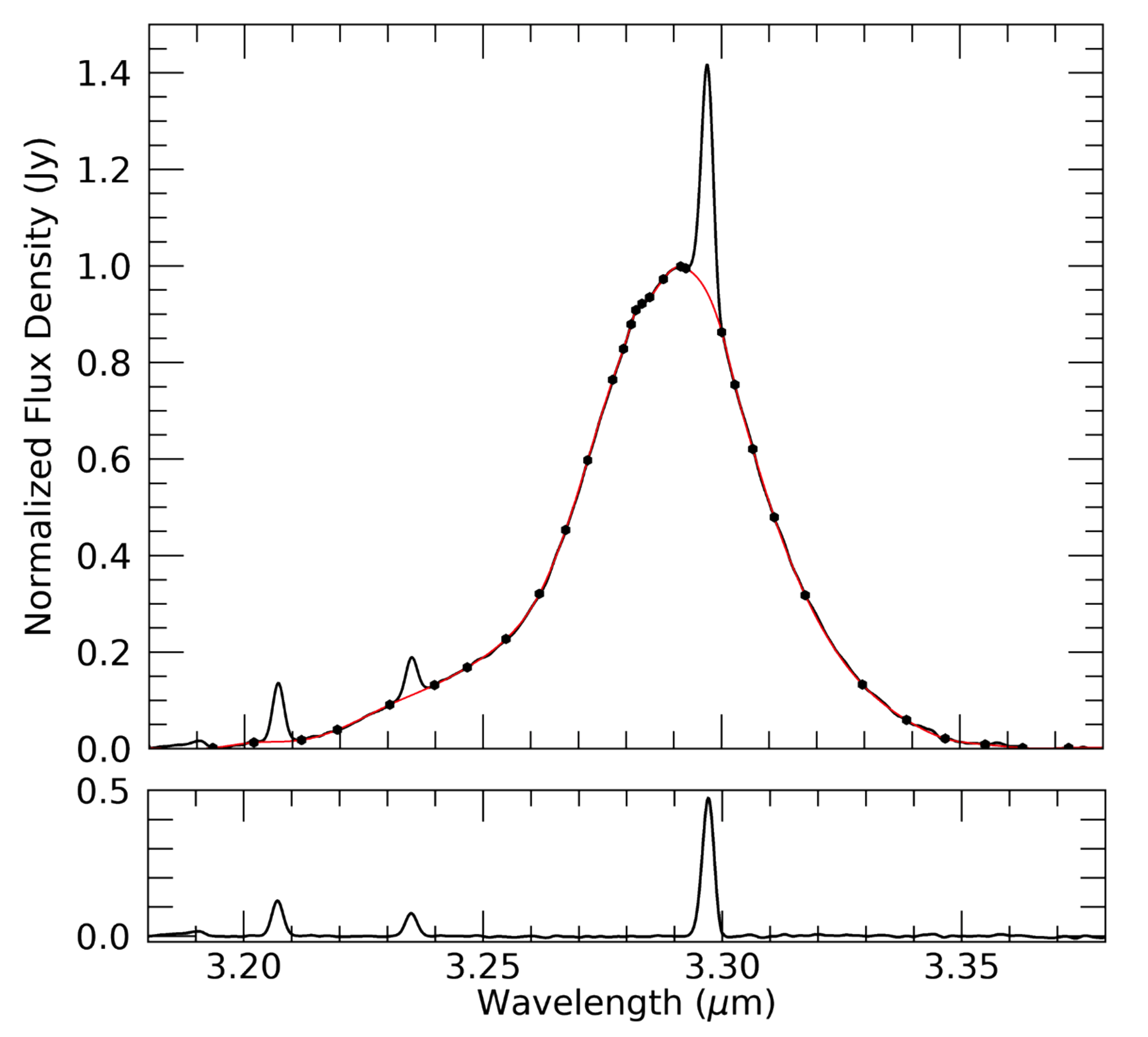}
%% original file "Fig 2 new.pdf"
\caption{(Top) Spline fit to the weighted average of the NGC 7027 IEF at 3.3 $\mu$m (black solid line). The points selected for the spline fit are shown as filled black dots and the spline fit as the red solid line. Note that the continuum and the plateau emission at 3.4 $\mu$m have been removed as described in the text.  The long-wavelength wing is affected by the assumed profile of the 3.4 $\mu$m plateau emission and is therefore uncertain. (Bottom) Difference of the 3.3 $\mu$m IEF and the spline fit.  Lines at 3.207, 3.234, and 3.297 are the \mbox{[\ion{Ca}{4}],} H$_2$ 1$\rightarrow$0 O(5), and H Pfund-$\delta$ lines, respectively. }  \label{fig: spline fit} 
\end{figure}  

We present a clean profile of the 3.3 µm IEF that will be useful for future comparisons to laboratory data of possible carriers of the emission band.  We first used a spline fit to the NGC 7027 weighted average spectrum. The spline fit and the residual of the fit is shown in Figure 2. We show in Figure 3 the 3.3 µm IEF profile of NGC 7027 without the emission lines. This profile is provided as an online table to aid future work on the 3.3 µm IEF, especially with the high quality spectra that will be available from JWST. {\em Any plausible carrier hypothesis must be able to closely match this profile (i.e., central wavelength, width, and asymmetry) and its invariance for highly diverse physical environments (e.g., circumstellar shells, \ion{H}{2} regions, reflection nebula, and the diffuse ISM).}  
 
The ISO spectra of M82 and NGC 253 show that the Class A emission profile and central wavelength also holds on galaxy wide spatial scales and this can also be seen in the AKARI spectra of extragalactic objects  \citep{Inami18, Doi19}. This shows that the Class A profile is ubiquitous and dominant in galaxies, and it highlights the invariance of the Class A 3.3 $\mu$m IEF.

There are less common 3.3 $\mu$m IEF profiles such as the Class B spectrum of HD 44179 \citep{Tokunaga91, Song03, Song07, vanDiedenhoven04} as well as the rare and unusual spectra of HD 97048 and Elias 1 \citep{vanKerckhoven02}. These special cases provide additional clues to the nature of the carrier material, especially the spatial variations in the 3.3 $\mu$m IEF observed in HD 44179 \citep[see][Figure 9]{Song07} but are not discussed in this paper.

%% The "ht!" tells LaTeX to put the figure "here" first, at the "top" next
%% and to override the normal way of calculating a float position
%% =========================== Figure 3
\begin{figure}[ht]
%% \epsscale{1.2}
\plotone{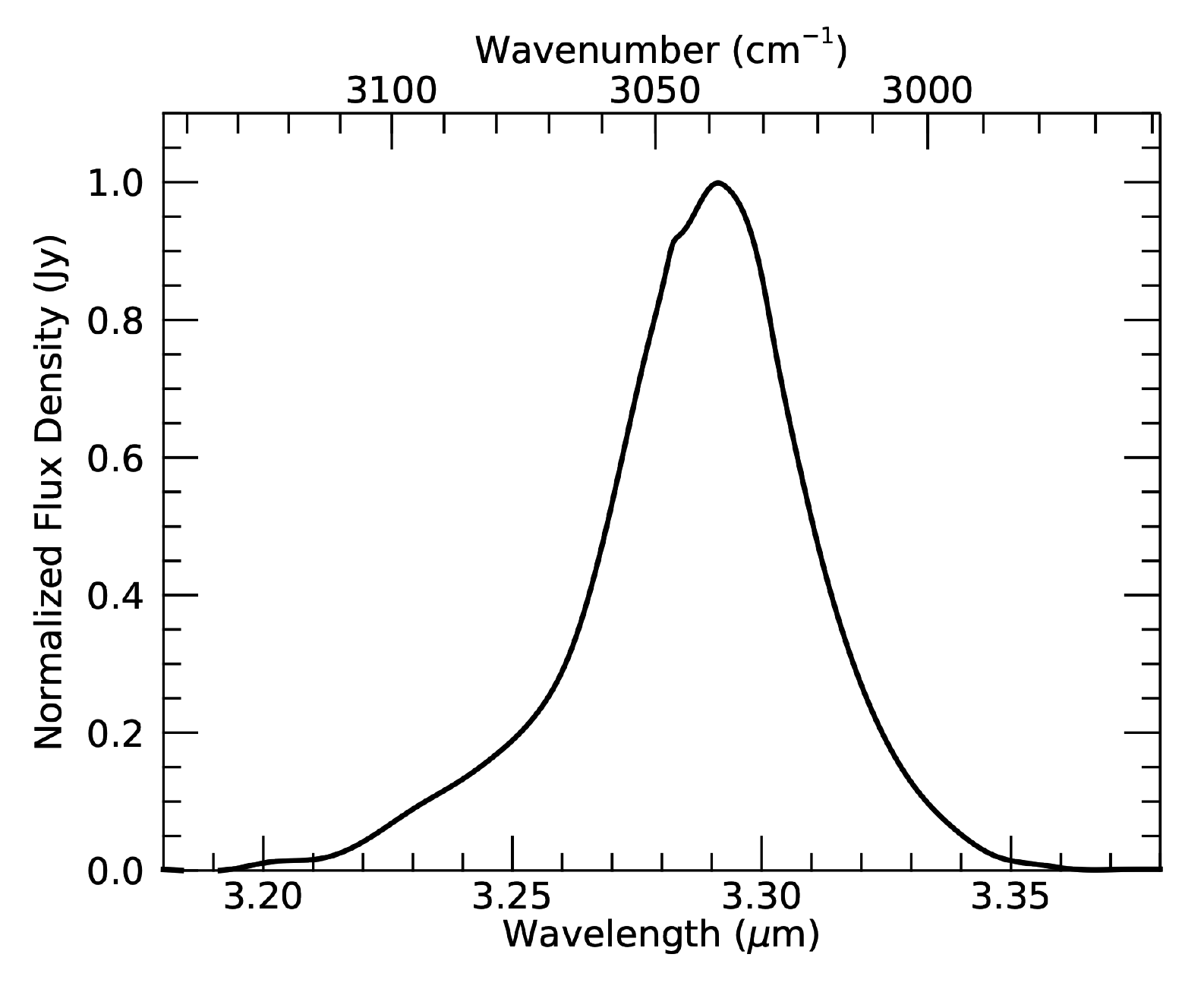}
%% original file "Fig 3 thickLine cut.pdf"
\caption{Spline fit to the weighted average of the NGC 7027 IEF at 3.3 $\mu$m.   The central wavelength and FWHM of this profile are 3.289 $\mu$m (3040 cm$^{-1}$) and 0.041 $\mu$m (38 cm$^{-1}$), respectively.   
The data points in this spectrum can be obtained from the data behind the table.
\label{fig: IEF profile}}
\end{figure}

\pagebreak
%% =============================================== Sec. 3
\section{Analysis of the Laboratory PAH spectra of the 3.3 $\mu$\lowercase{m} CH(S) Band} \label{sec:lab spectra}

\subsection{Fitting the Laboratory PAH Widths and Shifts}
%% Section 3.1

The gas phase laboratory measurements of \citet{Joblin94, Joblin95-1} provide high spectral resolution (1 cm$^{-1}$) spectra of the 3.3 $\mu$m band for different size PAHs over a temperature range of about 400 – 900 K.  We focus on the interpretation of their laboratory spectra for the 3.3 $\mu$m CH(S) band of naphthalene (C$_{10}$H$_8$), pyrene (C$_{16}$H$_{10}$), coronene (C$_{24}$H$_{12}$), and ovalene (C$_{32}$H$_{14}$).  The temperature and size dependent widths and central wavelengths for $N_C$$>$32 are difficult to obtain both experimentally and theoretically. Since there is general agreement that $N_C$$>$50 for the interstellar PAHs we  extrapolate the laboratory results to larger $N_C$.  We consider in Section 3.3 two extrapolation approaches that we argue represent plausible limits to the actual behavior.  The theoretical underpinnings of the extrapolation are discussed in Section 3.4. 

\citet{Joblin95-1} found that the widths and shifts for each PAH scaled linearly with temperature.  Thus, each PAH is characterized by two temperature slopes, one for the band width, and one for the band shift.  Here, we carry the analysis of \citet{Joblin95-1} a step further and show that each of these temperature slopes also scales linearly with $N_C$.  \cite{Pech02} did a similar analysis but they did not extrapolate to $N_C$$>$50. 

Our fits to the laboratory temperature-dependent slopes of the band widths and shifts of the center absorption frequencies, denoted as $\chi_w$ and $\chi_c$, are presented in Figure 4.  Both slopes are well represented by linear fits, suggesting that extrapolation to larger PAH molecules is meaningful.  In addition to the slopes, there is an intercept specific to each molecule that corresponds to the band width and peak frequency at a reference temperature, which will be discussed later.  For the smallest PAHs, like naphthalene, the width of its rotational envelope is approximately equal to its vibrational band width. This requires a correction to the observed width in order to compare to the observed band widths of the larger PAH molecules, for which the rotational envelope contribution is negligible (see Appendix A).

%% =========================== Figure 4
%% The "ht!" tells LaTeX to put the figure "here" first, at the "top" next
%% and to override the normal way of calculating a float position
\begin{figure}[htbp]
\plotone{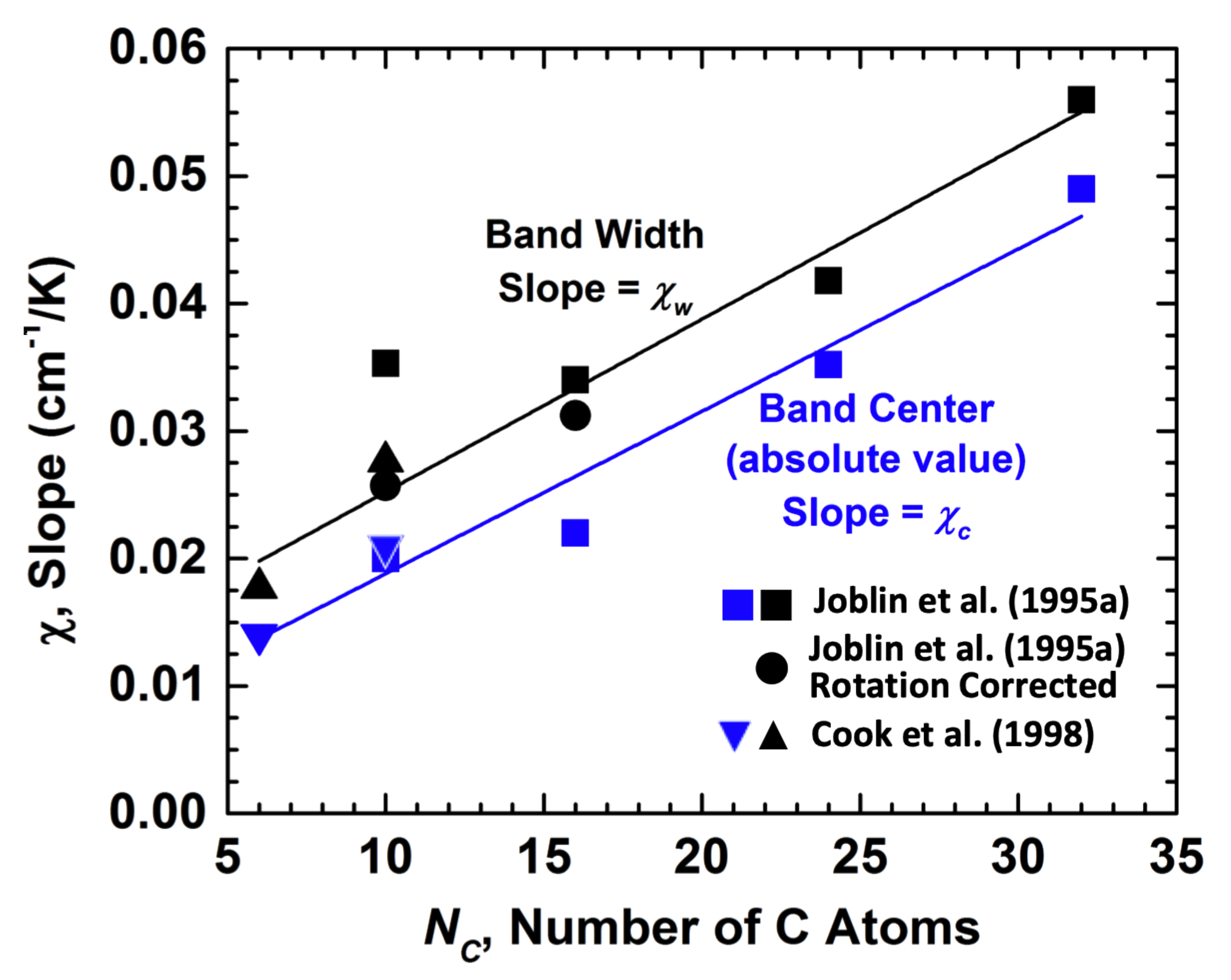}
%% original file "Fig 4 rev3.8 final.pdf"
\caption{Size dependence of the temperature slopes for the widths (black squares) and shifts of the center absorption frequencies (blue squares) for the 3.3 $\mu$m bands of naphthalene (C$_{10}$H$_8$), pyrene (C$_{16}$H$_{10}$), coronene (C$_{24}$H$_{12}$), and ovalene (C$_{32}$H$_{14}$) \citep{Joblin95-1}.  The solid lines are least squares linear fits to the rotationally corrected data of \citet{Joblin95-1} and the data of \citet{Cook98} discussed below.  Filled black circles for the $N_C$ = 10 and 16 points denotes the correction for the rotational envelope associated with naphthalene and pyrene (see Appendix A). The fit to the widths was performed using the corrected points.  Note that the shift of the band center frequency is negative, but for display purposes we show the absolute value.  The fits are given by $\chi_w$ = 0.0117 $+$ 0.00136$N_C$ for the width and $\chi_c$ = $-$0.00606 $-$ 0.00127$N_C$ for the band center frequency shift.  The standard deviation of the fits were $\pm$0.0018  and $\pm$0.0026  for $\chi_w$ and $\chi_c$ respectively. The filled blue and black triangles show the center shift and width slopes derived from the single UV photon excitation experiments of \citet{Cook98} (see text for details).}   \label{fig: slope vs Nc}
\end{figure}

Additionally, we have extended the lower size limit of \citet{Joblin95-1} to benzene \citep{Cook98}, such that the linear fits are demonstrated over a factor of approximately 5 in size (i.e. $N_C$ = 6--32).  This also suggests that a linear dependence may plausibly extend to $N_C$$>$32.  On the other hand, we recognize that a linear dependence may not be accurate for very large $N_C$, so we consider the possibility of an asymptotic limiting behavior for the slopes.  We define a conservative limit to the quantification of the asymptotic limit (see Section 3.3 for details).  {\em Most importantly, we note that the basic conclusions of this work that the widths and central wavelengths will vary significantly with variations of excitation energy and PAH size do not depend on precise knowledge of the behavior of the slopes for larger $N_C$.} 

\citet{Cook98} used UV laser single photon excitation of PAH molecules in the size range of $N_C$ = 6--24 to observe the resulting spectral emission of the 3.3 $\mu$m band.  The experimental conditions enabled high levels of internal excitation while maintaining low levels of rotational excitation.  In this respect, these measurements are a more realistic simulation of the interstellar environment than those of \citet{Joblin95-1}. For benzene, we used equations (1) and (2), presented in Section 3.3, to determine $\chi_w$ and $\chi_c$, respectively, based on the spectrum shown by \citet{Cook98} in their Figure 8.  A comparable approach was used for naphthalene based on their Figure 7 and the parameter values reported in their Table 13.  The results for $\chi_w$ represent an average based on the widths for the top three, highest vibrational energies, differenced with the lower energy width for $T$ = 930 K. 

Because benzene is a relatively small molecule it has relatively large rotational constants.  Even though the rotational temperature was low, around 300 K, the breadth of the rotational envelope ($\approx$24 cm$^{-1}$) is significant compared to the observed width of 73 cm$^{-1}$ for the 193 nm (6.4 eV) photon excitation.  The low measurement spectral resolution, 18 cm$^{-1}$, is also a significant source of broadening.  We retrieved the intrinsic vibrational broadening by deconvolving the rotational envelope and spectral resolution broadening to get $\gamma_{vib} = 66.5 = (73^2 - 18^2 - 24^2)^{1/2}$, which follows the example shown in Appendix A, Figure 7.

\subsection{Application of Laboratory PAH Spectra to Observational Spectra}
%% Subsection 3.2

There are a number of factors to consider in the application of laboratory data to modeling and analysis of the observational data.  The laboratory data is measured at discrete temperatures for which all the rotational and vibrational degrees of freedom are characterized by the gas kinetic temperature.  For the observational data, each rotational and vibrational degree of freedom may be characterized by different effective temperatures.  As discussed below, the spectral broadening due to the rotational envelope is not substantial for PAHs with $N_C$ greater than approximately 20.  
Thus, for interstellar PAHs where $N_C >$50, the rotational envelope broadening is negligible.  However, we do need to remove the effect of rotational envelope broadening for the smaller PAHs measured in the laboratory, since we are seeking to quantify the purely vibrational contributions for the widths and shifts.  

It is commonly assumed, as we do here, that for a fixed internal energy there is a single effective vibrational temperature that characterizes all the modes.  An observational spectrum arises from the absorption of a UV photon followed by fast intramolecular energy transfer and then an IR emission cascade within the ground electronic state.  This means that an observational spectrum is a superposition of a continuous distribution of spectra, each corresponding to a different internal energy, or equivalently, to a different internal vibrational temperature as the molecule cools.

Laboratory emission spectra at different temperatures can be summed to simulate a cascade spectrum.  A significant advantage to using laboratory emission data is that it correctly represents the complexities of Fermi resonances, which are difficult to accurately model with theoretical methods, particularly for large PAHs.  If only temperature-dependent absorption spectra are available, they can be converted to emission spectra by multiplying each absorption spectrum by the Planck blackbody function at the appropriate temperature.  The methodology for calculating a cascade spectrum using an ensemble of temperature-dependent measured or simulated laboratory spectra is well established \citep[i.e.][]{Leger84, Joblin95-1}.  Our goal here is to modify the laboratory widths and shifts for the effects of the IR cascade.

We are looking to associate a cascade width and shift to the UV excitation energy and PAH size.  The absorption of a UV photon results in an initial, maximum internal vibrational temperature, $T_{\rm max}$.  A method for determining $T_{\rm max}$ for a spectral distribution of UV photons from a stellar source with effective temperature $T_{\rm eff}$ is presented in the following section.  
Due to the radiative cooling, the temperatures that characterize the width and shift are lower than $T_{\rm max}$.  
We represent these temperatures using $\beta T_{\rm max}$ where $\beta$$<$1. As described in Section 3.3, we empirically estimated $\beta$ from cascade calculations presented in \citet{Joblin95-1}. Their cascade calculations were based on Lorentzian profiles tied to the temperature-dependent width and peak location of the experimental data. 

We expect that $\beta$ will be sensitive to the frequency of the vibrational fundamental.  If one is considering a relatively high frequency mode, such as a CH(S) mode, then a relatively small  decrease in temperature will significantly decrease the emission intensity.  Hence, only a small temperature interval will contribute to the cascade spectrum.  Thus, we anticipate $\beta$ will be close to 1.

We expect that $\beta$ will be different for the band profile width and shift.  
For the band center shift, the component spectral profiles shift increasingly to higher frequency as the cascade temperature decreases  \citep{Joblin95-1}. 
This means that the cascade shift will be less than the initial shift for $T_{\rm max}$, hence $\beta$$<$1.  

The cascade width is less straightforward to understand because it arises from the interplay of several factors.
Each component of the cascade spectrum shifts to higher frequency, thus extending higher frequency edge of the band, which suggests that $\beta$$>$1. However, both the width of each component spectrum and its overall contribution (i.e., weighting factor) decrease with the decreasing cascade temperature. 
The combined effect of these factors results in a cascade width that is
very close to the initial width for $T_{\rm max}$ 
(i.e., $\beta$$\approx$1).

\subsection{Modification of the Laboratory Widths and Shifts for a Radiative Cascade}  %% Section 3.3

Given the time-dependent cooling curve for a particular PAH, the width and peak frequency shifts in Figure 4 can be used to model the full radiative cascade spectrum as discussed by \citet{Joblin95-1}.  By scaling the calculated line profiles in Fig. 8 of \citet{Joblin95-1}, we derived a reasonably accurate (approximately $\pm$10\%), simple approximation to the width, $\Delta \nu$, of the cascade spectrum given by

\begin{equation}   %% Eq. (1)
\Delta  \nu = \Delta  \nu  \left( 500\ {\rm K} \right) + \chi_w \left(  \beta_w T_{\rm max }-500 \right) 
\left( 1-\exp  \left( -{N_S}/{N_C} \right)  \right) 
\quad\mbox{${\rm for}\ T_{\rm max} > 500{\rm K},$}
\end{equation}
%%\begin{multiline}   %% Eq. (1)
%%\Delta  \nu = \Delta  \nu  \left( 500K \right) + \chi_w \left(  \beta_w %%T_{\max }-500 \right) \times \\
%%\indent \left( 1-\exp  \left( -{N_S}/{N_C} \right)  \right) 
%%\quad\mbox{$for\ T_{max} > 500K$}\,\,\,\, (1)
%%\end{multiline}

\noindent where 500 K is a reference temperature,  $\chi_w$ is the band width slope as a function of $N_C$ (see Figure 4), $T_{\rm max}$ is the vibrational temperature corresponding to the energy of the absorbed photon, $\beta_w$ accounts for the modest reduction of $T_{\rm max}$ due to the radiative cascade, and $N_S$ is a PAH size associated with the asymptotic large molecule/{\em solid-state}  limit (see below).  The value of $T_{\rm max}$ is determined by integration of the molecule-specific heat capacity (Appendix B) until the integrated energy is equal to the absorbed photon energy, $E_{\rm UV}$.  Some examples of the relationship of $T_{\rm max}$ to $E_{\rm UV}$ for different size PAHs are shown in Table 2.  $\Delta \nu$(500 K) is molecule dependent and varies from $\approx$15--20 cm$^{-1}$ \citet{Joblin95-1}. For simplicity we adopted a value of $\Delta \nu$(500 K) = 20 cm$^{-1}$ because the majority of molecules (3 out of 4) investigated by \citet{Joblin95-1} exhibited values close to 20 cm$^{-1}$.  A value of  $\beta_w$ = 0.96 was derived from the computed cascade spectra in \citet{Joblin95-1} (see their Figure 8 for the 10 eV excitation spectra).

%% ==================================== Table 2
\begin{table}[h!]
\begin{center}
    \begin{minipage}{3.0in}
    \begin{center}
        \caption{Vibrational Temperatures $T_{\rm max}({\rm K})$ for Different PAH Sizes and the Absorbed UV Photon Excitation Energies ($E_{\rm UV})$} \label{vib temp}
        \begin{tabular}{ccccc}  \hline
        PAH / $E_{\rm UV}$ & 6 eV & 8 eV & 10 eV & 12 eV \\ \hline
        C$_{16}$H$_{10}$ & 1627 & 1976 & 2322 & 2661 \\
        C$_{32}$H$_{14}$ & 1096 & 1299 & 1498 & 1692 \\
        C$_{66}$H$_{20}$ & 762  & 882  & 996  & 1106 \\
        C$_{96}$H$_{24}$ & 647  & 739  & 829  & 912  \\ \hline
        \end{tabular} 
    \end{center}
\end{minipage}
\end{center}
\end{table}
	
Analogous to the widths, we use the relationships and cascade calculations in Fig. 8 of \citet{Joblin95-1} to derive an approximate relationship for the band center frequency, $\nu_c$, to PAH size and its associated excitation temperature.  This relationship is given by

\begin{equation}   %% Eq. (2)
\nu_c = \nu_c \left( 0\ {\rm K} \right) + \chi_c \left(  \beta_c T_{\rm max }-200 \right) 
\left( 1-\exp  \left( - {N_S}/{N_C} \right)  \right)  \quad\mbox{${\rm for}\ T_{\rm max} > 200{\rm K},$} 
\end{equation}
%%\begin{multiline}   %% Eq. (2)
%%\nu_c = \nu_c \left( 0K \right) + \chi_c \left(  \beta_c T_{\max }-200 %%\right) \times \\
%%\indent \left( 1-\exp  \left( - {N_S}/{N_C} \right)  \right)  %%\quad\mbox{$for\ T_{max} > 200K$} \,\,\,\, (2)
%%\end{multiline}

\noindent where $\chi_c$ is the band center slope as a function of $N_C$ (see Figure 4) and $\nu_c$(0 K) is the fundamental frequency for a selected PAH.  We note that $\chi_c$ exhibits a very tight linear relationship to $N_C$, suggesting that the extrapolation to larger PAH is reasonable.  While the absolute band shift, $\nu_c$, is molecule specific through $\nu_c$(0 K) \citep{Joblin95-1}, we consider the relative band shift, $\Delta \nu_c$ = $\nu_c - \nu_c$(0 K), which depends principally on the PAH size via $T_{\rm max}$ and $N_C$ in equation (2).  As for $\beta_w$, a value of $\beta_c$ = 0.87 was determined from the laboratory data of \citet{Joblin95-1}.  

\citet{Pech02} suggested that for very large PAH molecules the width and shift slopes should reach asymptotic values, which they refer to as the solid-state limit.  We have introduced the solid-state size parameter $N_S$ into equations (1) and (2) to account for the asymptotic behavior as a function of $N_C$.  The value of this parameter is not known;
however, we can estimate a lower limit by tying the quickest onset of the solid-state limit to the uncertainty in the linear fits of the width and shift with respect to $N_C$.  
A value of $N_S$ = 103 is obtained from the exponential term in Eqs. (1) and (2), from which $N_S$ = $- N_C$ ln(0.04) = 103, where $N_C$ = 32 and the average uncertainty of the two linear fits is 0.04.  

\subsection{Theoretical Rationale for the Extrapolation Approximations}
%% subsection 3.4

\citet{Joblin95-1} derived a theory-based expression (eq.(6) in their paper) for the temperature-dependent band shift based on the assumption that that the shifts were primarily due to the anharmonic coupling terms, $\chi_{ij}$, between a C--H stretch band and the much lower frequency vibrational modes ($\nu_{max}$ $\ll$ 700 cm$^{-1}$).  They demonstrated that for a given PAH, the band shift scales linearly with the vibrational temperature.  We reexpress their formula to show that it also supports a linear scaling with PAH size.  The modified expression is

\begin{equation}  %% Eq. (3)
  \Delta  \nu_{c,i}= \left( 3N_{C}-6 \right) \frac{kT}{hc} \sum _{j=1}^{j_{\rm max }}\frac{f_{j} \langle \chi_{i,j} \rangle }{ \nu_{j}}  
\end{equation}

\noindent where $\Delta \nu_{c,i}$ is the band center shift for the $i$th C--H stretch fundamental, 3$N_C$$-$6 is the total number of skeletal carbon vibrational modes, $\nu_j$ is the mean frequency for the low frequency modes in the $j$th frequency bin (i.e., we are binning the modes into frequency intervals), $f_j$ is the fraction of the total modes in the $j$th bin,  $\langle \chi_{ij} \rangle$ the average anharmonic coupling constant for all the modes in the $j$th bin, and $j_{\rm max}$ corresponds to the high frequency cutoff, $\nu_{\rm max}$.  

Assuming the same frequency bins for different size PAHs, and the same $\nu_{max}$, we empirically determine using the PAHdb that $f_j$ for compact PAHs does not depend on PAH size.  Thus, the only size dependence in the summation terms must arise from the 
$\langle \chi_{ij} \rangle$ term. The observational fact that $\Delta \nu_{c,i}$ varies linearly with $N_C$ implies that $\langle \chi_{ij} \rangle$ is size independent up to $N_C$=32. Since there is no physical reason to expect a strong discontinuity in the scaling beyond $N_C$=32 it is not unreasonable to assume that the linear scaling with $N_C$ persists for larger $N_C$.  It seems plausible to expect that 
$\langle \chi_{ij} \rangle$ may gradually decrease for larger $N_C$, which provides the motivation for introducing a solid-state limit.

The assumption that $f_j$ is independent of PAH size for the low frequency C skeletal modes of compact PAHs is tested as follows. Consider the compact PAHs coronene, C$_{24}$H$_{12}$, ovalene, C$_{32}$H$_{14}$, circumovalene, C$_{66}$H$_{20}$, and circumcircumcoronene, C$_{96}$H$_{24}$, which have 3$N_C$-6 vibrational degrees of freedom of 66, 90, 192 and 282, respectively.  According to Joblin et al. (1995) the anharmonicity arises primarily from modes with frequencies $\ll$700 cm$^{-1}$.  We assume an upper limit frequency of 350 cm$^{-1}$ and find from the NASA PAH database that the number of modes below this limit are 9, 16, 29, and 43 for C$_{24}$H$_{12}$, C$_{32}$H$_{14}$, C$_{66}$H$_{20}$, and C$_{96}$H$_{24}$, respectively.  The corresponding $f$ factors are $9/66=0.14$, $16/90=0.18$, $29/192=0.15$, and $43/282=0.15$. There is good consistency of the $f$ factors over a substantial range of PAH sizes. 

We note that the five laboratory PAHs in Figure 4 are relevant exemplars of the proposed interstellar PAHs in that they are examples of highly symmetric compact PAHs.  This offers some confidence that extrapolation of their properties to larger PAHs of the same symmetry type is reasonable.  \citet{Ricca12} have shown that a small number of highly symmetric, compact PAHs in the $N_C$=60--100 size range may be the dominant carriers for the Class A profiles. 

The preceding discussions pertains solely to the band center shifts.  We do not have a comparably simple expression for the band widths.  However, the observational fact that the band widths correlate tightly with the band shifts (the absolute values of the slopes are nearly identical) suggests the use of the same extrapolation assumptions as for the band center shifts.

\subsection{Computational Chemistry for Determination of Widths and Shifts}
%% subsection 3.5

Theoretical methods for calculating anharmonicity effects in PAH molecules are under active development \citep{Mackie18, Mulas18, Peeters21}.  The theoretical calculations are challenging and are not yet sufficiently accurate to substitute for experimental measurements.  For example, using the band profiles calculated by \citet[Supplemental Material]{Mackie18}  we estimate the vibrational width slopes, $\chi_w$, for naphthalene and pyrene to be 0.057 and 0.045 cm$^{-1}$ $\rm K^{-1}$, respectively.  These estimates are based on the excitation temperatures and widths for the 3 and 5 eV cascade spectra.  For comparison, the experimental width slopes, with rotational broadening removed, for naphthalene and pyrene are 0.026 and 0.031 cm$^{-1}$ $\rm K^{-1}$, respectively.  The theoretical predictions are high by roughly a factor of 1.5--2.0 and they decrease with increasing PAH size, $N_C$.

%% =========================== Figure 5
%% The "ht!" tells LaTeX to put the figure "here" first, at the "top" next
%% and to override the normal way of calculating a float position
\begin{figure}[b!]
%% \epsscale{1.23}
\plotone{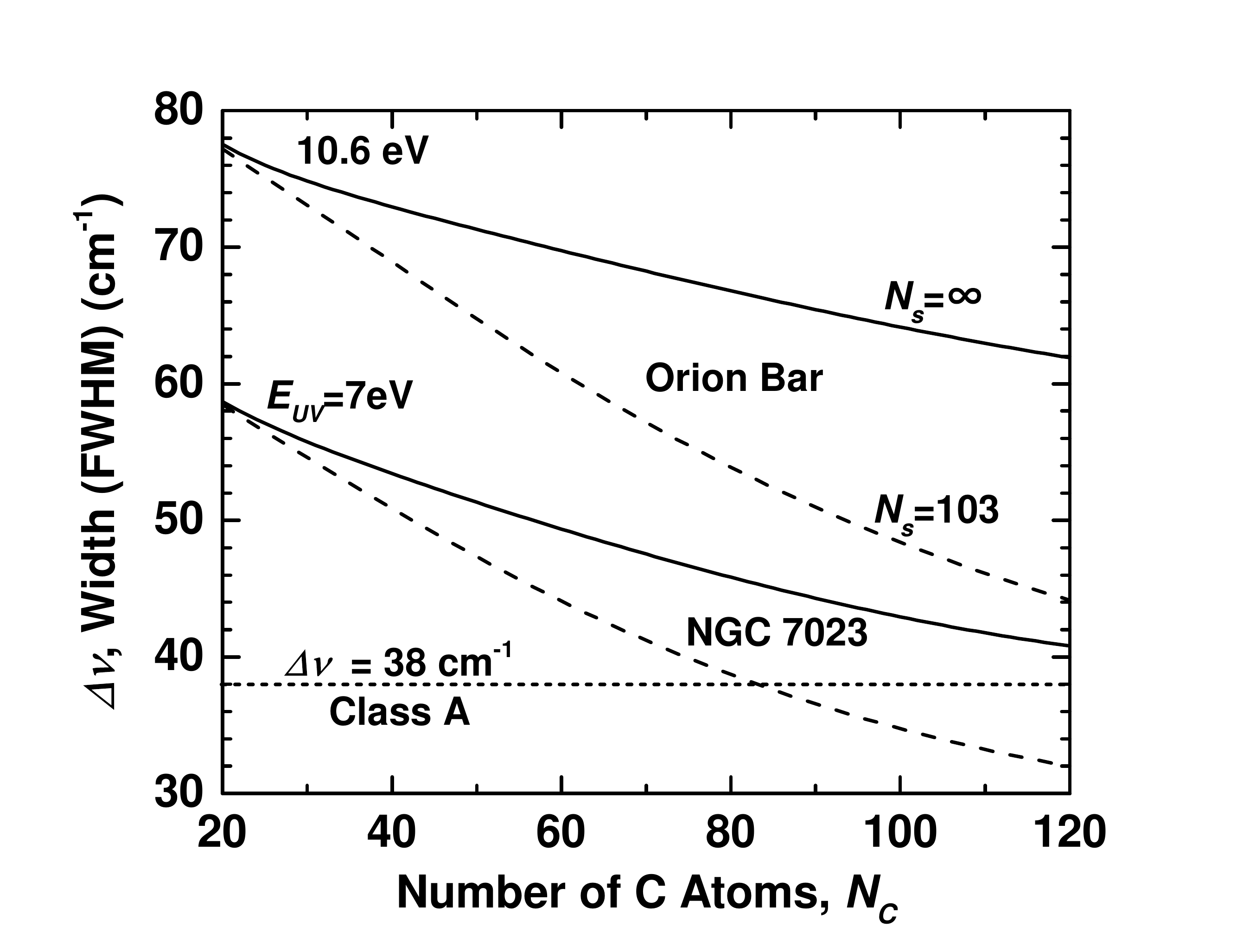}
%% original file "Fig 5 rev3.2 final.pdf"
\caption{Predicted widths of the CH(S) band for different size PAH molecules at UV excitation energies of 7 and 10.6 eV based on Equation (1).  The characteristic width for a Class A 3.3 $\mu$m IEF band, $\Delta \nu$ = 38 cm$^{-1}$, is indicated by the dashed lines.  
There are two curves for each excitation energy that correspond to different values for the solid-state limit, $N_S$, defined in Section 3.3.  $N_S$ = $\infty$ corresponds to linear width and shift slopes for all PAH sizes.  $N_S$ = 103 corresponds to the quickest onset of the solid-state PAH size limit consistent with the uncertainty of the straight line fit to the band width slope in Figure 4.      \label{fig: widths}}
\end{figure}

\section{Implications for PAHs in the ISM}
%% Section 4.

\subsection{Widths and Shifts of PAHs in the ISM}
%% subsection 4.1

We used the scaling relationships given in Equations (1) and (2) to predict the widths and shifts in the ISM, accounting for the radiative cooling of the molecule, for a range of PAH sizes and for two UV photon excitation energies.  The predictions are displayed in Figures 5 and 6 for excitation energies of 7.0 and 10.6 eV. 
For comparison, we indicate the observed width of 38 cm$^{-1}$ for the ubiquitous Class A profile in Figure 5 and the band center frequency shift of $-$30 cm$^{-1}$ in Figure 6.

%% =========================== Figure 6
%% The "ht!" tells LaTeX to put the figure "here" first, at the "top" next
%% and to override the normal way of calculating a float position
\begin{figure}[t!]
%% \epsscale{1.23}
\plotone{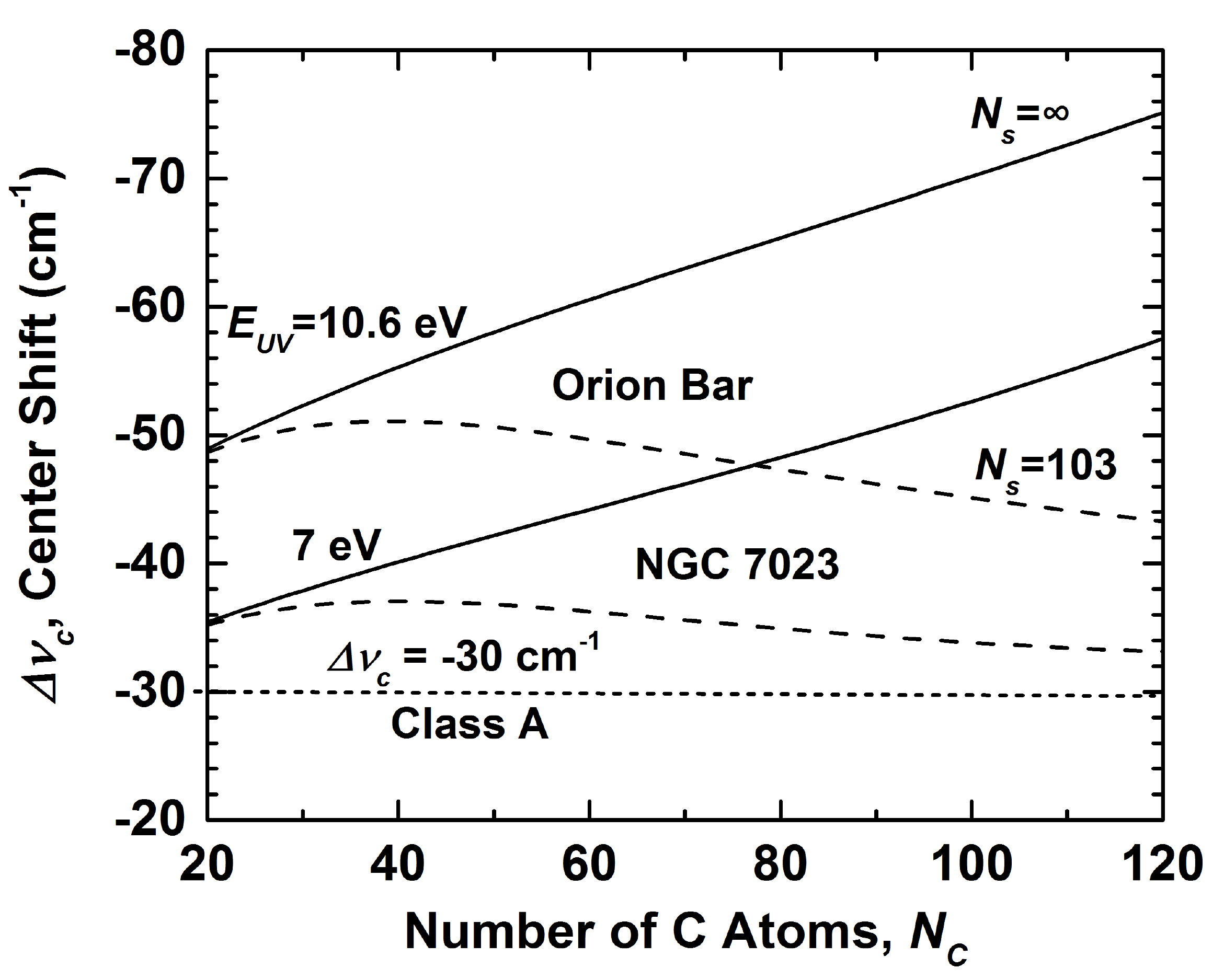}
%% original file "Fig 6 rev3.2 final.pdf"
\caption{Predicted band center shifts for different size PAH molecules at UV excitation energies of 7 and 10.6 eV relative to the zero-point frequency (the calculated fundamental frequency) of the CH(S) band. The predicted shift of the Class A 3.3 $\mu$m IEF assumes that the zero-temperature band frequencies are the same for the compact PAHs studied by \citet{Joblin95-1} and is not applicable to all PAHs.  
This assumption is supported by density functional theory (DFT) calculations, which show that the zero-temperature fundamental band frequencies is nearly constant for $N_C$ $>$ 20 where $\Delta \nu_c$(0 K) $\approx$ 3070 cm$^{-1}$ from the PAHdb.  
The observed shift for the Class A  3.3 $\mu$m band, $\Delta \nu$ = $-$30 cm$^{-1}$, is indicated by the short dashed line.  
There are two curves for each excitation energy that correspond to different values for the solid-state limit, $N_S$, defined in Section 3.3.  }
    \label{fig: shift} 
\end{figure}

We selected $E_{\rm UV}$ = 7.0 and 10.6 eV because they are representative of low and high average UV photon absorption energies for PAHs in different environments.  
For simplicity, we frame our analysis around the widths and shifts based on a single profile at the average excitation energy, $<$$E_{\rm UV}$$>$.  The distribution of excitation energies is broad and the average of the profiles over $E_{\rm UV}$ may not produce a width and shift identical to those for a single profile evaluated at 
$E_{\rm UV}$.  However, we are focused on the general trends with respect to PAH size and excitation energy, which should be insensitive to the method used to calculate the widths and shifts. 

In order to associate a particular stellar UV source and its effective  temperature, $T_{\rm eff}$, with an average UV excitation energy, $<$$E_{\rm UV}$$>$, we derived an empirical relationship between $<$$E_{\rm UV}$$>$ and $T_{\rm eff}$ based on the points from \citet{Andrews15}, where ($<$$E_{\rm UV}$$>$, $T_{\rm eff}$) is (6.2 eV, 14,000 K) for NGC 1333, (7.0 eV, 17,000 K) for NGC 7023, and (8.1 eV, 23,000 K) for NGC 2023.  The relationship is

\vspace{-0.2in}
\begin{equation}         %% Eq. (4)
    <E_{\rm UV}> \, = \, 12.3 \, \{ 1- \exp  \left( -4.9 \times 10^{-5} \, T_{\rm eff} \right)  \}
\end{equation}

\noindent where  $<$$E_{\rm UV}$$>$ = 12.3 eV is an asymptotic limit at high $T_{\rm eff}$.  For $T_{\rm eff}$=40,000 K, which is characteristic of the Orion Bar \citep{Salgado16}, we obtain $<$$E_{\rm UV}$$>$=10.6 eV.

The $<$$E_{\rm UV}$$>$ derived above from $T_{\rm eff}$ represent an upper limit to the local excitation energy for a PAH carrier.  This is due to the spatially variable extinction from dust, which preferentially absorbs far UV compared to near UV photons \citep{Draine03}.  Consequently, as the stellar UV photons travel further from their source,  $<$$E_{\rm UV}$$>$ decreases.  Furthermore, assuming that the 3.3 $\mu$m IEF arises primarily from neutral PAHs, the neutrals will populate regions with relatively {\em low} values of $<$$E_{\rm UV}$$>$ where ionization does not occur.  Model predictions from Montillaud et al. (2013) show (see their Figure 2) that, for NGC 7023, the neutral PAHs favor regions further from the stellar source, with higher UV extinction (i.e. lower $<$$E_{UV}$$>$), than PAH cations.  

These considerations may contribute to an explanation of the invariance of the 3.3 $\mu$m band. It would require that the dust extinction for neutral PAHs always results in a low and relatively constant excitation energy. 
In general, the UV excitation spectrum of a PAH is bimodal (for example, see Figure 2 in \cite{Pech02}).  
The excitation spectrum is bimodal because the PAH absorption spectrum is bimodal with centroids around ~6 and ~12 eV \citep{Malloci04} and there is 
a  cutoff in the stellar spectrum at 13.6 eV due to the onset of H atom absorption). 
Thus, enhanced relative extinction of the 12 eV excitation could result in a source invariant excitation energy peak of approximately 6 eV. 

We infer from Figures 5 and 6 that neutral PAHs with $E_{\rm UV} \approx 6$ eV could produce shifts and widths consistent with observations of the 3.3 $\mu$m IEF for a distribution of PAHs with an average size of $<$$N_C$$>\approx 70$.  The preceding arguments may explain a constant excitation energy.
However, it is also necessary to explain the profile invariance which implies that the relative abundances of the PAHs remain invariant among diverse sources.

\subsection{Previous modeling studies of the 3.3 $\mu$m IEF}
%% subsection 4.2

The general characteristics (e.g., size, shape, chemical composition, and charge state) of the PAH carriers associated with the 3.3 $\mu$m IEF band are inferred from spectral fitting studies employing the PAHdb \citep[i.e.][]{Bauschlicher08, Bauschlicher09, Ricca12, Boersma18}. The PAHdb software tool computes the band positions and takes into account the full radiative cascade to the ground state. In order to account for the vibrational excitation energy and anharmonic effects a bandwidth of 15 cm$^{-1}$ and a frequency shift to lower frequency of 15 cm$^{-1}$ is typically assumed.  This constant bandwidth and frequency shift are applied irrespective of UV radiation field and PAH size.  This fitting procedure is at odds with the laboratory studies, such as those of  \citet{Joblin95-1} which show that the shifts and widths depend on the PAH size and the excitation energy. Furthermore, the laboratory shifts and widths are different for every band.  

Given that the 3.3 $\mu$m IEF is due to aromatic C--H stretching modes, then there must be accompanying IEFs due to out-of-plane (OOP) C--H bending modes.  One of the main IEF bands occurs at 11.2 $\mu$m and is ascribed to the OOP mode (Peeters 2011). Like the 3.3 $\mu$m IEF, the majority of observed 11.2 $\mu$m emission profiles are also of Class A.  The 3.3 and 11.2 $\mu$m bands are thought to arise primarily from neutral PAH molecules (Allamandola et al. 1999; Tielens 2008).  For neutral PAHs, these two features are the dominant features. In contrast, PAH cations produce most of their emission in several features spanning the 5--10 $\mu$m region. One of the strengths of the PAH model is that it can quantitatively account for the observed ratios of the integrated fluxes for these bands for different sources as well as the spatial variability of the ratio within a source \citep[i.e.][]{Schutte93, Croiset16, Maragkoudakis20}.  

Several studies have included some laboratory-based constraints in their IEF spectral models \citep{Verstraete01, Pech02}.  While many of the details differ between the two models, they share key assumptions central to achieving a nearly invariant 3.3 $\mu$m spectral profile for very different environments.  One is that the band width slope is fixed at that for coronene and does not vary with PAH size.  A second is a power-law size distribution, $N_C^{-3.5}$, with a sharp turn-on at a minimum size, $N_{C \rm min}$. The final assumption is that $N_{C \rm min}$ varies with variations in $<$$E_{\rm UV}$$>$ in such a manner as to preserve a constant excitation temperature.  Since the initial excitation temperature depends on the ratio $<$$E_{\rm UV}$$>$/$N_C$ \citep{Berne12}, keeping this ratio constant leads to a constant excitation temperature.  

There are several problems with these assumptions. First is that the results of this study strongly suggest that the band width slope varies with PAH size.  Second, the values of $N_{C \rm min}$ determined for the sources considered in these studies vary over the range of 20$-$30 C atoms.  This is problematic because PAHs with $N_C$ less than $\approx$50 should be fully dehydrogenated \citep{Allain96-1, Montillaud13, Andrews16, Joblin20}.  
The relatively fast size distribution falloff combined with the decrease in excitation temperature with increasing size means that only a small range of sizes, anchored by the minimum size, contribute to the modeled spectrum.  For example, we estimate that relative to $N_{C \rm min}$ = 50 the contribution of $N_{C}$ = 55 is decreased by about 50\%.

%% =============================== Sec. 5
\section{Summary} \label{sec:summary}

We explored the consequences of the invariance of the Class A 3.3 $\mu$m IEF under a wide range of physical conditions.  The 3.3 $\mu$m IEF band center, band width, and band profile are properties that are interconnected, requiring that a definitive carrier identification should match all these properties {\em simultaneously}.  

The key points of our paper are:

\begin{enumerate}

\item Extrapolation of the laboratory data:  Laboratory data on compact PAHs \citep{Joblin95-1, Chakraborty19} show that the band center wavelength and FWHM depends on temperature and the number of carbon atoms, $N_C$.  We use the laboratory data to extrapolate to larger PAH sizes that are thought to be representative of the PAHs sizes in the ISM.  

\item Fitting the 3.3 $\mu$m IEF center wavelength and width (FWHM): The observations show that the band center wavelength and width are constant for the Class A sources.  Combined with the extrapolated laboratory data, this implies that for a given excitation energy the PAH size is determined.  This suggests that the carrier species have a very narrow range of sizes with invariant relative abundance and are excited by the same average UV photon energy– otherwise one would see significant source to source profile variability.

\item Fitting the 3.3 $\mu$m IEF profile: The lack of spectral diversity in the band position, band width, and profile of the 3.3 $\mu$m IEF is difficult to explain (Sec. 4).  The identical Class A band profile for most galactic sources and the integrated light of galaxies implies a common carrier with comparable excitation in many galactic environments.

\end{enumerate}

Our conclusions rely on the extrapolation of the laboratory data to large $N_C$.  In view of the very strong constraints imposed on the size and excitation of the PAHs given the invariance of the 3.3 $\mu$m IEF, we highlight the critical need for high temperature gas phase spectra of PAHs with $N_C$ $>$ 50.  Although we are aware of the formidable difficulties of obtaining such spectra, the spectra would have a fundamental impact on our understanding of PAHs in the ISM.

We note that even if our extrapolation of the laboratory data is proven to be invalid in the future, the invariance of the 3.3 $\mu$m IEF under many excitation environments and evolutionary history challenges our understanding of the IEF carrier.  It will be extremely interesting to see if James Webb Space Telescope observations can show that the Class A 3.3 $\mu$m IEF is the same in the early universe as in the present epoch or equivalently, if low metallicity environments produce the same carrier.

%% ------------------- double anonymous --------------------

\acknowledgments

L.S.B. thanks R. Shroll and J. Quenneville of Spectral Sciences, Inc. for technical discussions.
We thank C. Boersma for help with the Ames PAHdb and G. Sloan for help with the ISO SWS spectral archive.
E. Peeters provided spectra from her database of ISO spectra.
We thank R. Knacke, K. Sellgren, T. Onaka, and an anonymous reviewer for helpful comments.
This research has made use of NASA’s Astrophysics Data System and the SIMBAD database, 
operated at CDS, Strasbourg, France.

%% ------------------- double anonymous --------------------


\vspace{5mm}
\facilities{ISO(SWS)}


\software{NASA Ames PAH IR Spectroscopic Database \citep{Bauschlicher18}}

%% APPENDIX =============================

\appendix

%% \pagebreak

%% ======================== Appendix A

\section{Removal of the Rotational Broadening Envelope from the Naphthalene 3.3 $\mu$\lowercase{m} Laboratory Spectrum}

The observed band widths for the gas phase PAHs investigated in the laboratory by \citet{Joblin95-1} are comprised of two main components: (1) the width due to the rotational envelope for a single vibrational band and (2) the width due to vibrational anharmonicity, which arises from the overlap of many shifted single band rotational envelopes.  For naphthalene, unlike the larger PAH molecules investigated by \citet{Joblin95-1}, its rotational envelope makes a substantial contribution to the overall band width.  In order to make a consistent comparison to the anharmonicity-dominated widths of the larger PAHs, we need to remove the rotational envelope contribution to the total band width.    

The relative contribution of the rotational envelope decreases rapidly with increasing PAH size.  The rotational envelope width is proportional to 1/$N_C$ while the vibrational anharmonicity width is proportional to $N_C$; hence, the ratio of the rotational envelope width to the  vibrational width decreases rapidly as 1/$N_C^2$.  As described below, removal of the rotational contribution from the observed profile requires spectral deconvolution, as opposed to simply subtracting the rotational envelope width. 

%% =========================== Figure 7
%% The "ht!" tells LaTeX to put the figure "here" first, at the "top" next
%% and to override the normal way of calculating a float position
\begin{figure}[b!]
\plotone{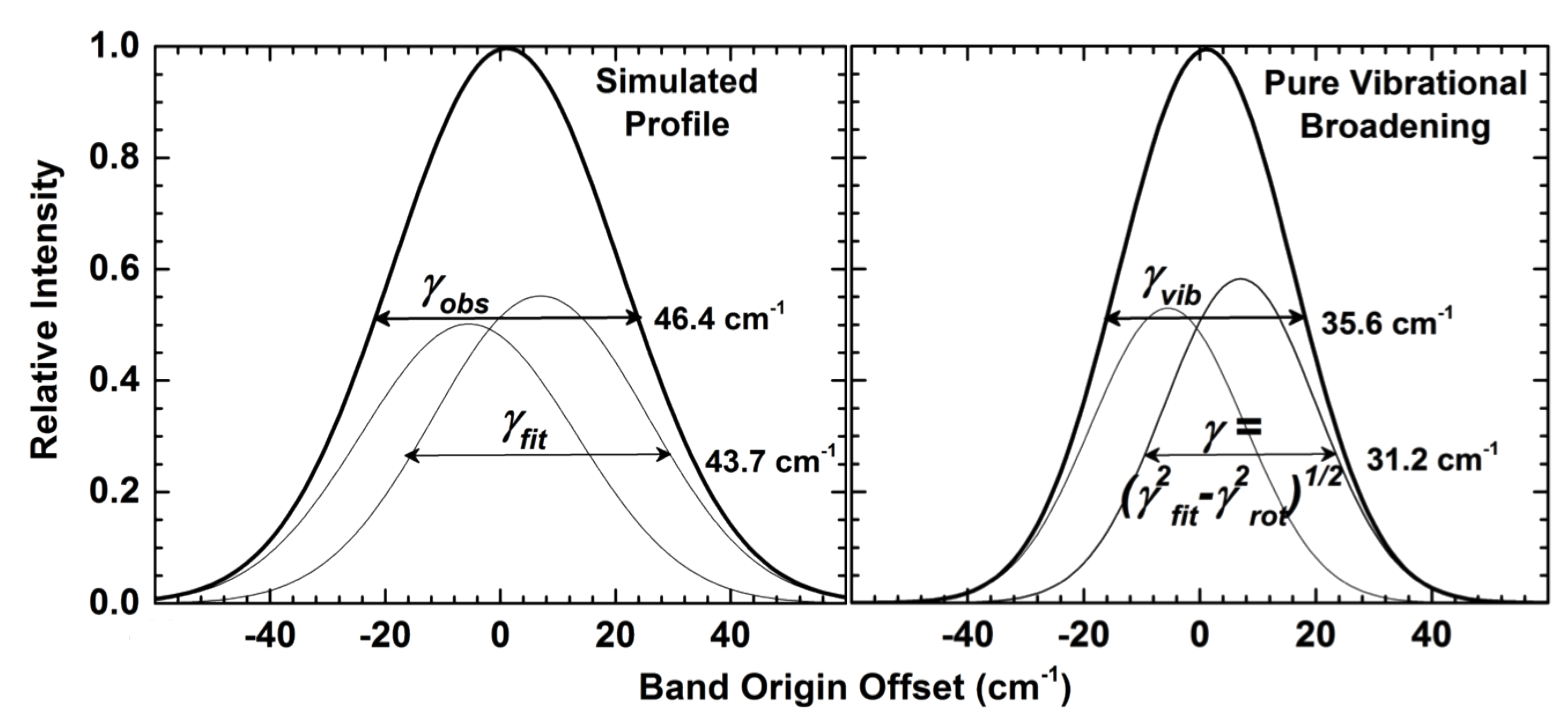}
%% original file "Fig 7 new.pdf"
\caption{(Left panel) Simplified model for the 3.3 $\mu$m laboratory naphthalene emission spectrum at 800 K (thick black line).  The profiles for the two contributing fundamental bands (thin black lines) are shown and are summed together to get the observed emission band profile.  The two bands are separated by 14 cm$^{-1}$.  The identical widths for the two fundamental bands, $\gamma_{\rm fit}$, were adjusted until the width of the composite spectrum matched the width of the observed laboratory spectrum, $\gamma_{\rm obs}$.  (right panel) The profile for the purely vibrationally broadened composite spectrum (thick black line), obtained by summing the two components from which the rotational broadening, $\gamma_{\rm rot}$, was removed from the fitted component bands, 
$\gamma_{\rm fit}$ (see formula shown in the figure).  The width of the resulting pure vibrational broadened line is $\gamma_{\rm vib}$, and this quantity is used to determine the corrected data point for naphthalene in Fig. 4.     \label{fig: sim profile}}
\end{figure}

To determine the band width slope parameter $\chi_w$, we employed an approximate model of the contributions of the temperature-dependent rotational and vibrational broadening to the naphthalene spectrum.  The model components are illustrated in Figure 7.  The naphthalene 3.3 $\mu$m band is composed of two fundamental bands of nearly equal intensity, separated by 14 cm$^{-1}$ (NASA Ames PAHdb; Bauschlicher et al. 2018).  The vibrational and rotational spectral distributions for each fundamental band are represented by separate temperature-dependent Gaussian profiles.  When both broadening mechanisms apply, the vibrational and rotational contributions are convolved (i.e., the widths are not added).  The temperature-dependent rotational widths, 
$\gamma_{\rm rot}$,  were estimated from rotational profile calculations using the PGOPHER code (Western 2017) that are based on the rotational constants for naphthalene ($A = 0.106$ cm$^{-1}$ and $(B + C)/2 = 0.035$ cm$^{-1}$).  For simplicity, we modeled naphthalene as a symmetric top molecule  and the rotational width is the approximate width of the rotational envelope comprising of the P, Q, and R branches.  At each temperature, we determined the vibrational width that, when convolved with the rotational width, produced a total band width consistent with the laboratory spectra.  By determining the vibrational width, $\gamma_{\rm vib}$, at two temperatures, we defined the band width slope parameter using $\chi_w = ( \gamma_{\rm vib}(T_1)- \gamma_{\rm vib}(T_2))/(T_1 - T_2)$.  The result is $\chi_w$ = 0.026 cm$^{-1}$ $\rm K^{-1}$, which is a factor of 0.74 lower than the observed slope of 0.035 cm$^{-1}$ $\rm K^{-1}$.  The correction for the rotational broadening of pyrene is considerably smaller, a factor of 0.96.  The corrections for the larger PAH molecules, coronene and ovalene are negligible.

%% ======================== Appendix B

\section{Heat Capacity Approximation for PAHs}

For convenience, we used a simplified approximation to the vibrational heat capacity.  The exact heat capacity equation for a nonlinear molecule is given by equation 7 in  \citet{Leger89},

\begin{equation}
    C_v/k = \sum_{i=1}^{3N-6}\frac{ \theta_{i}^2 \, \exp ( - \theta_i ) }{ ( \exp  ( - \theta_i ) - 1 ) ^2}
\end{equation}

\noindent where $\theta_i = h \nu_i / kT$ for the $i$th vibrational mode, and $N$ is the total number of atoms in the molecule.  We approximated the heat capacity using a two-mode expression,

\begin{equation}
    C_v/k = \sum_{i=1}^{2} g_{i}  \frac{ \theta_{i}^2 \, \exp ( - \theta_i ) } { ( \exp ( - \theta_i ) - 1 ) ^2}
\end{equation}

\noindent where $g_i$  is a degeneracy factor.  The two effective modes correspond to the high frequency CH(S) band and to all the other lower frequency modes.  This results in $g =  N_{\rm H}$ for the CH(S) modes, where $N_{\rm H}$ is the number H atoms in the molecule, and $g =  (2 N_{\rm H} + 3N_{\rm C} - 6)$ for the lower frequency modes.  The two mode frequencies are $\nu$ = 3075 and 881 cm$^{-1}$ for the CH(S) mode and low frequency modes, respectively.  The low frequency mode was adjusted to provide a good fit to the more rigorous heat capacity calculations of \citet{Joblin95-1}.  The accuracy of the simplified method is within $\approx$2\% of the exact results.

%% ========= References =============

\bibliography{references}{}
\bibliographystyle{aasjournal}

%\allauthors.  


\end{document}